\documentclass[12pt]{JHEP3}

\usepackage{cite}
\usepackage{enumerate}
\usepackage{amsbsy}
\usepackage[dvips]{graphicx}
\usepackage{graphics}

\newcommand{\be}{\begin{equation}} \newcommand{\ee}{\end{equation}}
\newcommand{\bea}{\begin{eqnarray}} \newcommand{\eea}{\end{eqnarray}}

\newcommand{\re}[1]{(\ref{#1})}

\newcommand{\pat}{\partial}

\renewcommand{\sec}[1]{section \ref{#1}}
\newcommand{\fig}[1]{figure \ref{#1}}
\newcommand{\brt}[1]{[#1]}

\renewcommand{\a}{\alpha}
\renewcommand{\b}{\beta}

\newcommand{\LCDM}{$\Lambda$CDM\ }
\newcommand{\GN}{G_{\mathrm{N}}}
\newcommand{\ha}{\frac{1}{2}}

\newcommand{\keq}{k_{\mathrm{eq}}}
\newcommand{\aeq}{a_{\mathrm{eq}}}
\newcommand{\teq}{t_{\mathrm{eq}}}

\newcommand{\rmd}{\mathrm{d}}

\newcommand{\bz}{\bar{z}}

\newcommand{\nonum}{\\}
\newcommand{\etal} {et al.}

\newcommand{\adot}{\dot{a}}
\newcommand{\addot}{\ddot{a}}
\newcommand{\rhodot}{\dot{\rho}}

\newcommand{\HH}{\frac{\adot^2}{a^2}}

\newcommand{\av}[1]{\langle{#1}\rangle}
\newcommand{\sQ}{\mathcal{Q}}
\newcommand{\sR}{{^{(3)}R}}

\newcommand{\om}{\omega_{\mathrm{m}}}

\newcommand{\PRD}[1]{{\it Phys. Rev.} {\bf D#1}}

\newcommand{\PRL}[1]{{\it Phys. Rev. Lett.} {\bf #1}}

\newcommand{\PLA}[1]{{\it Phys. Lett.} {\bf A#1}}

\newcommand{\MNRAS}[1]{{\it Mon. Not. Roy. Astron. Soc.} {\bf #1}}
\newcommand{\APJ}[1]{{\it Astrophys. J.} {\bf #1}}

\newcommand{\JHEP}[1]{{\it JHEP} {\bf #1}}
\newcommand{\CQG}[1]{{\it Class. Quant. Grav.} {\bf #1}}
\newcommand{\GRG}[1]{{\it Gen. Rel. Grav.} {\bf #1}}
\newcommand{\AaA}[1]{{\it Astron. \& Astrophys.} {\bf #1}}
\newcommand{\PROG}[1]{{\it Prog. Theor. Phys.} {\bf #1}}

\newcommand{\IJMPD}[1]{{\it Int. J. Mod. Phys.} {\bf D#1}}

\title{Backreaction as an alternative to dark energy and modified gravity}

\author{Syksy R\"{a}s\"{a}nen\thanks{Contribution to the proceedings
    of the ``Beyond the Concordance Model'' workshop held at The Stellenbosch Institute for Advanced Study, 23.-27. August 2010.} \\

University of Helsinki, Department of Physics \\
P.O. Box 64, FIN-00014 University of Helsinki, Finland \\

\email{syksy {\it dot} rasanen {\it at} iki {\it dot} fi}}

\abstract{
The predictions of homogeneous and isotropic cosmological
models with ordinary matter and gravity are off by a factor
of two in the late universe. One possible explanation is the
known breakdown of homogeneity and isotropy
due to the formation of non-linear structures.
We review how inhomogeneities affect the average expansion
rate and can lead to acceleration,
and consider a semi-realistic model where the observed
timescale of ten billion years emerges from structure formation.
We also discuss the relation between the average expansion
rate and observed quantities.
}

\preprint{HIP-2010-34/TH}

\begin{document}

\setcounter{tocdepth}{2}

\setcounter{secnumdepth}{3}

\section{Introduction} \label{sec:intro}

\subsection{A factor of 2}

The early universe, at least from Big Bang Nucleosynthesis at one
second onwards, is well described by a model which is exactly
homogeneous and isotropic (up to linear perturbations) and
contains only ordinary matter, and where the relation between
matter and spacetime geometry is given by ordinary general
relativity. Here ordinary matter means that the pressure is
non-negative, and ordinary general relativity is
based on the four-dimensional Einstein-Hilbert action.
Such a model also works well when applied to the late universe,
except that the distance and the expansion rate are
underpredicted by a factor of two.
The observed distance to the last scattering surface at redshift
1090 is a factor of 1.4--1.7 longer than predicted by the
spatially flat homogeneous and isotropic model dominated by
pressureless matter, keeping the Hubble parameter today fixed
(and assuming a power-law spectrum of primordial perturbations)
\cite{Vonlanthen:2010}.
Observations of type Ia supernovae and large-scale structure
are consistent with this cosmic microwave background (CMB)
measurement, and they show that the discrepancy arises
at redshifts of order one and smaller, when the universe is
about ten billion years old.
In the homogeneous and isotropic Friedmann-Robertson-Walker
(FRW) models, the explanation for the longer distances
is simple: the expansion of the universe has accelerated,
so objects have receded further away.

Most cosmological observations probe distances, but there
are also some measurements of the expansion rate.
Galaxy ages \cite{ages} and the radial baryon acoustic
oscillation signal \cite{BAOradial}
have been used to measure the expansion rate as a
function of redshift, and the value of the Hubble parameter
today is known with some accuracy \cite{Hubble}.
The expansion rate observations agree well with the distance
observations and support the interpretation of faster expansion
being the cause of the longer distances.
The measured Hubble parameter is about a factor of 2 larger
than expected compared to the matter density,
$\Omega_\mathrm{m0}\equiv8\pi\GN\rho_\mathrm{m0}/(3H_0^2)\approx0.25$
\cite{Peebles:2004},
or a factor of 1.2--1.5 larger if compared to the age
of the universe, $H_0t_0\approx$ 0.8--1
instead of $H_0t_0=2/3$ \cite{Krauss:2003}.

As the observations are beyond reasonable doubt, at least
one of the three assumptions of the theoretical model is wrong.
Either there is exotic matter with negative pressure (dark
energy), general relativity does not hold on cosmological scales,
or the homogeneous and isotropic approximation is not valid
at late times.

Apart from observations of the expansion rate and the
distance scale, there is no evidence for negative-pressure
matter or modifications of general relativity.
For example, such matter has not been detected in the laboratory,
nor have deviations from general relativity been observed in the
Solar system\footnote{Apart from possibly the Pioneer anomaly
and the flyby anomaly \cite{solar}.}.
Likewise, no objects have been seen to accelerate away
from each other. All of the relevant observations involve averages
over large volumes or integrals over large distances.
The situation is different from that of dark matter, for which
there is evidence from several kinds of observations
on various scales and eras, including local physics,
such as the CMB peak structure,
large-scale structure, rotation curves of galaxies, the
motions of galaxies and gas in clusters, gravitational
lensing and so on \cite{Roos:2010}.
This is the reason why constructing alternatives to dark matter
requires resort to baroque models, if it is possible at all \cite{teves}.
In contrast, the various observations usually interpreted
as indicating dark energy or modified gravity are all
different tracers of the same quantity: longer distances
and faster expansion.
An effect which leads to faster expansion and correspondingly
increases the distances can explain all of these observations.
(Indeed, the only effect of the favorite dark energy candidate,
vacuum energy, is to change the expansion rate.)

Unlike the assumptions of ordinary matter and gravity, the
assumption of only linear deviations from homogeneity and
isotropy is known to be violated at late times due to the
formation of non-linear structures, a process which has
a preferred time of about ten billion years.
Before concluding that new physics is needed to explain
the observations, we should therefore study the possibility
that the factor two failure of the predictions of the simple
homogeneous and isotropic models is due to the known
breakdown of homogeneity and isotropy
\cite{Buchert:2000, Tatekawa:2001, Wetterich:2001, Schwarz:2002, Rasanen, Kolb:2004}.

\subsection{Our clumpy universe}

It is important to distinguish between {\it exact} homogeneity
and isotropy and {\it statistical} homogeneity and isotropy.
Exact homogeneity and isotropy means that the space has a
local symmetry: all points and all directions are equivalent.
Statistical homogeneity and isotropy simply means that if we
consider a box anywhere in the universe, the mean quantities
in the box do not depend on its location, orientation or size,
provided that it is larger than the homogeneity scale.

The early universe is nearly exactly
homogeneous and isotropic, in two ways.
First, the amplitude of the perturbations around homogeneity
and isotropy is small.
Second, the distribution of the perturbations is statistically
homogeneous and isotropic. At late times, when density perturbations
become non-linear, the universe is no longer locally near homogeneity
and isotropy, and there are deviations of order unity in quantities
such as the local expansion rate. However, the distribution of the
non-linear regions remains statistically homogeneous and isotropic
on large scales. The homogeneity scale appears to be around
100 Mpc today \cite{Hogg:2005} (though see \cite{inhom}).

Due to the statistical symmetry, the average expansion rate evaluated
inside each box is equal (up to statistical fluctuations), but this
does not mean that it would be the same as in a completely smooth
spacetime, because there are structures in the box. The feature that
the average evolution of a clumpy space is not the same as the
evolution of a smooth space is called {\it backreaction}
\cite{Ellis:2005, Rasanen:2006b, Buchert:2007}.
We can say that time evolution and averaging do not commute:
if we smooth a clumpy distribution and calculate the time
evolution of the smooth quantities with the Einstein equation,
the result is not the same as if we evolved the full clumpy
distribution and took the average at the end.
Put simply, FRW models describe universes
which are exactly homogeneous and isotropic. They do not
describe universes which are only statistically homogeneous
and isotropic.
The effect of clumpiness on the average was first discussed
in detail by George Ellis in 1983 under the name
{\it fitting problem} \cite{fitting}.
Clumpiness affects the expansion of the universe, the way light
propagates in the universe and the relationship between the two.
The possibility that the late time deviations from the
simple homogeneous and isotropic models would be explained
in terms of these changes due to structure formation can
be called the {\it backreaction conjecture}.

In section 2 we go through the basics of how structures affect the
expansion rate. An increased average expansion rate due clumpiness
may be a bit unfamiliar concept, so in section 3 we explain what
average acceleration means physically using a simple toy model,
and in section 4 we go on to discuss a semi-realistic model for the
universe where we can get some numbers out. We also
briefly mention the relation of the backreaction
problem to the fascinating question of the Newtonian limit
of general relativity.
In section 5 we discuss the relation of the
average expansion rate to light propagation.
The order is a bit backwards, as it is the observed redshifts
and distances which are the important quantities, but it is
perhaps easier to start from the expansion rate.
We conclude in section 6 with a summary.

\section{Backreaction, exactly}

\subsection{The local expansion rate}

We consider a universe where the energy density of
matter dominates over the pressure, anisotropic stress
and energy flux everywhere. In other words, the matter
can be considered a pressureless ideal fluid, or dust.
We also assume that the relation between the matter and
the geometry is given by the Einstein equation:
\bea \label{Einstein}
  G_{\a\b} &=& 8\pi\GN T_{\a\b} = 8\pi\GN \rho u_{\a} u_{\b} \ ,
\eea

\noindent where $G_{\alpha\beta}$ is the Einstein tensor, $\GN$ is
Newton's constant, $T_{\alpha\beta}$ is the energy--momentum tensor,
$\rho$ is the energy density and $u^{\alpha}$
is the velocity of observers comoving with the dust.

The evolution and constraint equations can be written elegantly
in terms of the gradient of $u_\a$ and the electric and magnetic
components of the Weyl tensor \cite{Tsagas:2007},
\bea \label{gradu}
  \nabla_\b u_\a
  &=& \frac{1}{3} h_{\a\b} \theta + \sigma_{\a\b} + \omega_{\a\b} \ ,
\eea

\noindent where $h_{\a\b}$ projects orthogonally to $u^\a$:
roughly speaking, if $u^\a$ is understood as the time direction,
then $h_{\a\b}$ spans the spatial directions\footnote{Strictly speaking,
if the vorticity does not vanish, there are no hypersurfaces
orthogonal to $u^\a$.}.
The trace $\theta\equiv\nabla_\a u^\a$ is the volume expansion rate,
the traceless symmetric part $\sigma_{\a\b}$ is the shear tensor
and the antisymmetric part $\omega_{\a\b}$ is the vorticity tensor.
For an infinitesimal fluid element, $\theta$ indicates how
its volume changes in time, keeping the shape and the orientation
fixed, while the shear changes the shape and vorticity changes the
orientation. Roughly speaking, $\theta$ is the time derivative
of the volume element divided by the volume element.
In the FRW case, the volume expansion rate is just
$3 H$, where $H$ is the Hubble parameter.

The equations can be be decomposed into scalar, vector and tensor
parts with respect to the spatial directions orthogonal to $u^\a$.
We need only the scalar parts (we omit a scalar equation related
to the vorticity, which we don't need),
\bea
  \label{Rayloc} \dot{\theta} + \frac{1}{3} \theta^2 &=& - 4 \pi \GN \rho - 2 \sigma^2 + 2 \omega^2 \\
  \label{Hamloc} \frac{1}{3} \theta^2 &=& 8 \pi \GN \rho - \frac{1}{2} \sR + \sigma^2 - \omega^2 \\
  \label{consloc} \rhodot + \theta\rho &=& 0 \ ,
\eea

\noindent where a dot stands for derivative with respect to
proper time $t$ measured by observers comoving with the dust, 
$\sigma^2\equiv\ha\sigma^{\alpha\beta}\sigma_{\alpha\beta}\geq0$
and $\omega^2\equiv\ha\omega^{\alpha\beta}\omega_{\alpha\beta}\geq0$
are the shear and the vorticity scalar, 
respectively, and in the irrotational case, $\sR$ is the
spatial curvature of the hypersurface orthogonal to $u^\a$;
see \cite{Ellis:1990} for the definition in the case of
non-vanishing vorticity.

The equation \re{consloc} shows that the energy density is
proportional to the inverse of the volume, in other words
that mass is conserved. The second equation \re{Hamloc}
is the local equivalent of the Friedmann equation, and it relates
the expansion rate to the energy density, spatial curvature,
shear and vorticity.
The left-hand side of \re{Rayloc} gives the local acceleration.
Let us assume that the fluid is irrotational, i.e. that the vorticity
is zero. Taking into account vorticity would lead to technical
complications which we want to avoid \cite{Rasanen:2009b}.
As vorticity contributes positively to the acceleration,
putting it to zero gives a lower bound on the acceleration.
With no vorticity, the local acceleration is always negative,
or at most zero. This is just an expression of the fact that gravity
is attractive for matter satisfying the strong energy condition.
One might think that if we are interested in accelerated
expansion\footnote{Note that inferring from the observed
distances that the expansion has accelerated depends on assuming
the FRW relation between distance and the expansion rate. If
backreaction is important, this relation is modified; see \sec{sec:light}.
Based on direct measurements of the expansion rate, we can only say
that there has been less deceleration, not that the expansion has accelerated.},
there is no point in proceeding further. If the local expansion
rate decelerates everywhere and at all times, surely the average
expansion rate will also decelerate? However, this conclusion is false.

\subsection{The average expansion rate}

Let us consider the average expansion rate. The spatial
average of a scalar quantity $f$ is simply the integral over
the spatial hypersurface orthogonal to $u^\a$ (this is also
the hypersurface of constant proper time $t$ measured by the observers),
with the correct volume element, divided by the volume:
\bea \label{av}
  \av{f}(t) \equiv \frac{ \int d^3 x \sqrt{^{(3)}g(t,\bar{x})} \, f(t,\bar{x}) }{ \int d^3 x \sqrt{^{(3)}g(t,\bar{x})} } \ ,
\eea

\noindent where $^{(3)}g$ is the determinant of the metric on the
hypersurface of constant proper time $t$.

Averaging \re{Rayloc}--\re{consloc}, we obtain the
Buchert equations \cite{Buchert:1999}
\bea
  \label{Ray} 3 \frac{\addot}{a} &=& - 4 \pi \GN \av{\rho} + \sQ \\
  \label{Ham} 3 \HH &=& 8 \pi \GN \av{\rho} - \frac{1}{2}\av{\sR} - \frac{1}{2}\sQ \\
  \label{cons} && \pat_t \av{\rho} + 3 \frac{\adot}{a} \av{\rho} = 0 \ ,
\eea

\noindent where the backreaction variable $\sQ$ contains the effect
of inhomogeneity and anisotropy,
\bea \label{Q}
  \sQ \equiv \frac{2}{3}\left( \av{\theta^2} - \av{\theta}^2 \right) - 2 \av{\sigma^2} \ ,
\eea

\noindent and the scale factor $a(t)$ is defined so that the
volume of the spatial hypersurface is proportional to $a(t)^3$,
\bea \label{a}
  a(t) \equiv \left( \frac{ \int d^3 x \sqrt{ ^{(3)}g(t,\bar{x})} }{ \int d^3 x \sqrt{ ^{(3)}g(t_0,\bar{x})} } \right)^{\frac{1}{3}} \ ,
\eea

\noindent where $a$ has been normalised to unity at time $t_0$,
which we take to be today. As $\theta$ gives the expansion rate
of the volume, this definition of $a$ is equivalent to
$3\adot/a\equiv\av{\theta}$. We also use the notation $H\equiv\adot/a$.

The equations \re{Ray}--\re{cons} were first derived by
Thomas Buchert in 1999 \cite{Buchert:1999}.
They have a slightly different physical interpretation than
the FRW equations due to the different meaning of the scale factor.
In FRW models, the scale factor is a component of the metric,
and indicates how the space is evolving locally.
In the present context, $a(t)$ does not describe local
behaviour, and it is not part of the metric.
It simply gives the total volume of a region.

Mathematically, the Buchert equations differ
from the FRW equations by the presence of the
backreaction variable $\sQ$ and the related
feature that the average spatial curvature can have
non-trivial evolution.
In the FRW case, $\sQ=0$, and it then follows from the
integrability condition between \re{Ray} and \re{Ham}
that $\av{\sR}\propto a^{-2}$.
In general, $\sQ$ is non-zero, and it expresses
the non-commutativity of time evolution and averaging.
The backreaction variable $\sQ$  has two parts: the second term
in \re{Q} is the average of the shear scalar, which is also
present in the local equations \re{Rayloc}--\re{consloc}.
It is always negative (unless the spacetime is FRW,
in which case it is zero), and acts to decelerate the expansion.
In contrast, the first term in \re{Q}, the variance of
the expansion rate, has no local counterpart.
It may be called emergent in the sense that
it is purely a property of the average system.
The variance is always positive (unless the expansion
is homogeneous, in which case it is zero).
If the variance is sufficiently large compared to the
shear and the energy density, the average expansion rate
accelerates according to \re{Ray}, even though \re{Rayloc}
shows that the local expansion rate decelerates everywhere.

\section{Understanding acceleration} \label{sec:toy}

\subsection{A two-region toy model}

It may seem paradoxical that the average expansion rate accelerates
even though the local expansion rate slows down everywhere, so we first
look at a simple toy model to understand the physical meaning, before
venturing to a more realistic model.
Let us give the punchline right away.
In an inhomogeneous space, different regions expand at different
rates. Regions with faster expansion rate increase
their volume more rapidly, by definition.
Therefore the fraction of volume in faster expanding regions
rises, so the average expansion rate can rise.
Whether the average expansion rate actually does rise depends on
how rapidly the fraction of fast regions grows relative to the
rate at which their expansion rate is decreasing.

In the real universe, the initial distribution of density
(and thus of the expansion rate) is very smooth, with only
small local variations. In a simplified picture, overdense regions
slow down more as their density contrast grows, and eventually
they turn around and collapse to form stable structures.
Underdense regions become ever emptier, and their
deceleration decreases.
Regions thus become more differentiated, and the variance
of the expansion rate grows.

We can illustrate this with a simple toy model where there
are two spherically symmetric regions, one underdense
and one overdense \cite{Rasanen:2006a, Rasanen:2006b}.
We consider the regions to be Newtonian, so their evolution
is given by the spherical collapse model and the underdense
equivalent, i.e. they expand like dust FRW universes
with negative and positive spatial curvature, respectively.

We denote the scale factors of the underdense and the overdense
region by $a_1$ and $a_2$, respectively.
We take the underdense region, which models a cosmological void,
to be completely empty, so it expands like $a_1\propto t$.
The evolution of the overdense region, which models
the formation of a structure such as a cluster, is given by
$a_2\propto 1-\cos\phi$, $t\propto \phi-\sin\phi$, where
the parameter $\phi$ is called the development angle.
The value $\phi=0$ corresponds to the big bang singularity,
from which the overdense region expands until $\phi=\pi$,
when it turns around and starts collapsing. The region
shrinks to zero size at $\phi=2\pi$.
In studies of structure formation, the collapse
is usually taken to stabilise at $\phi=3\pi/2$
due to vorticity and velocity dispersion, and we also
follow the evolution only up to that point. The total
volume is $a^3=a_1^3+a_2^3$. The average expansion rate
and acceleration are
\bea
  \label{Hex} \!\!\!\!\!\!\!\!\!\!\!\!\!\!\!\!\!\!\!
H &=& \frac{ a_1^3 }{ a_1^3 + a_2^3 } H_1 + \frac{ a_2^3 }{ a_1^3 + a_2^3 } H_2 \equiv v_1 H_1 + v_2 H_2 \\
  \label{accex} \!\!\!\!\!\!\!\!\!\!\!\!\!\!\!\!\!\!\!
\frac{\addot}{a} &=& v_1 \frac{\addot_1}{a_1} + v_2 \frac{\addot_2}{a_2} + 2 v_1 v_2 (H_1-H_2)^2 \ .
\eea

The average expansion rate is the volume-weighted average of
the expansion rates $H_1$ and $H_2$, as one would expect.
It is therefore bounded from above by the fastest local expansion rate.
However, from the fact that both $H_1$ and $H_2$ decrease it does not
follow that their weighted average would decrease, or that the average
expansion rate would decelerate\footnote{Of course, an increase
in $H$ is not required for acceleration.}.
This is illustrated by the acceleration equation \re{accex}.
The first two terms are the volume-weighted average,
and because the regions decelerate (or at most have zero acceleration,
in the completely empty case), it is negative.
However, there is an additional term related to the difference between
the two expansion rates, which is always positive (as long as the
regions have non-zero volume and different expansion rates).
This term arises because a time derivative of \re{Hex}
operates not only on $H_1$ and $H_2$, but also on $v_1$ and $v_2$.
In terms of the general acceleration equation \re{Ray}, the first
two terms in \re{accex} come from the average density, and the
last term is the backreaction variable $\sQ$.

\begin{figure}
\hfill
\begin{minipage}[t!]{5.1cm} 
\scalebox{1.2}{\includegraphics[angle=0, clip=true, trim=0cm 2cm 0cm 0cm, width=\textwidth]{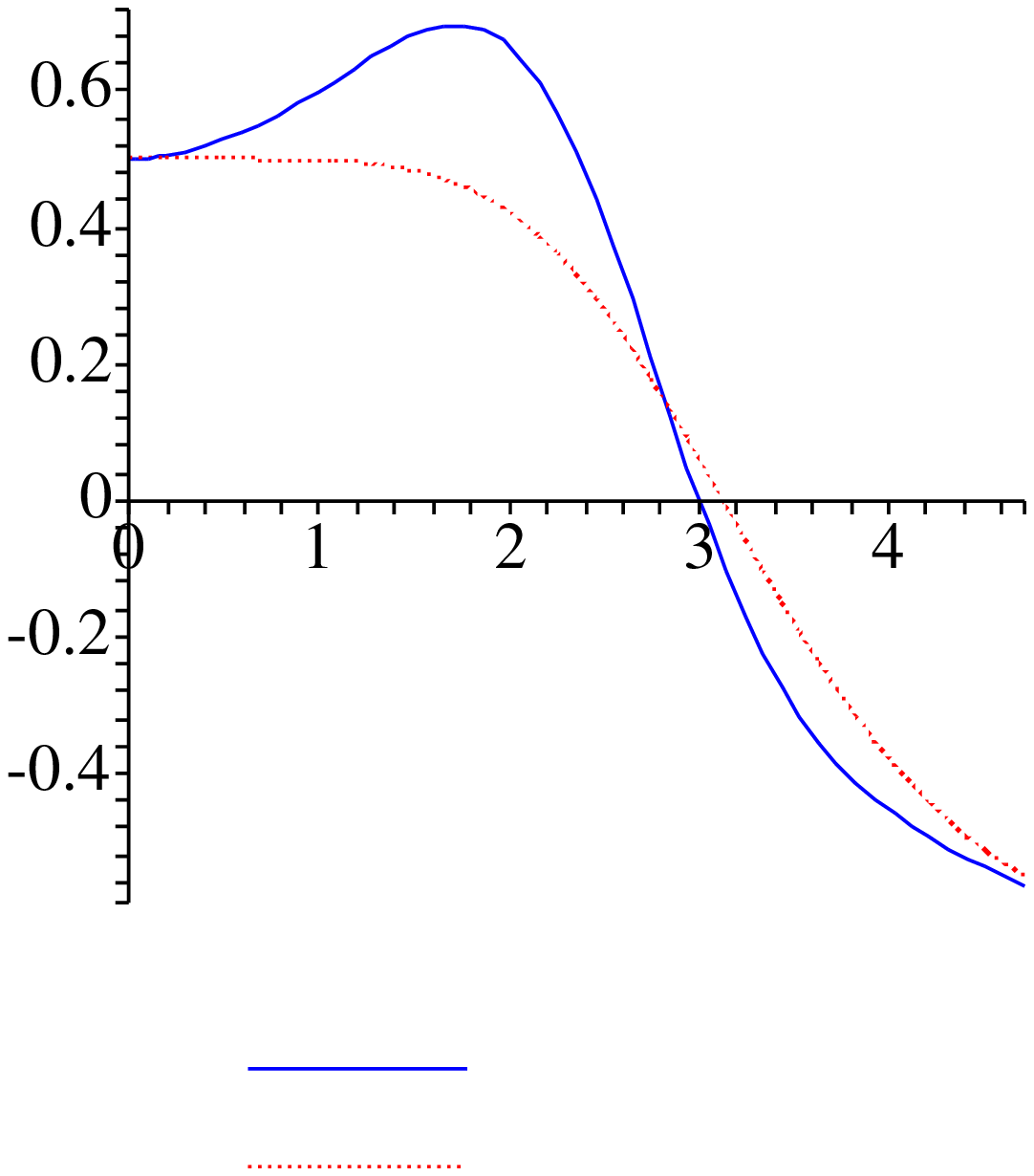}}
\begin{center} {\bf (a)} \end{center}
\end{minipage}
\hfill
\begin{minipage}[t!]{5.1cm}
\scalebox{1.2}{\includegraphics[angle=0, clip=true, trim=0cm 2cm 0cm 0cm, width=\textwidth]{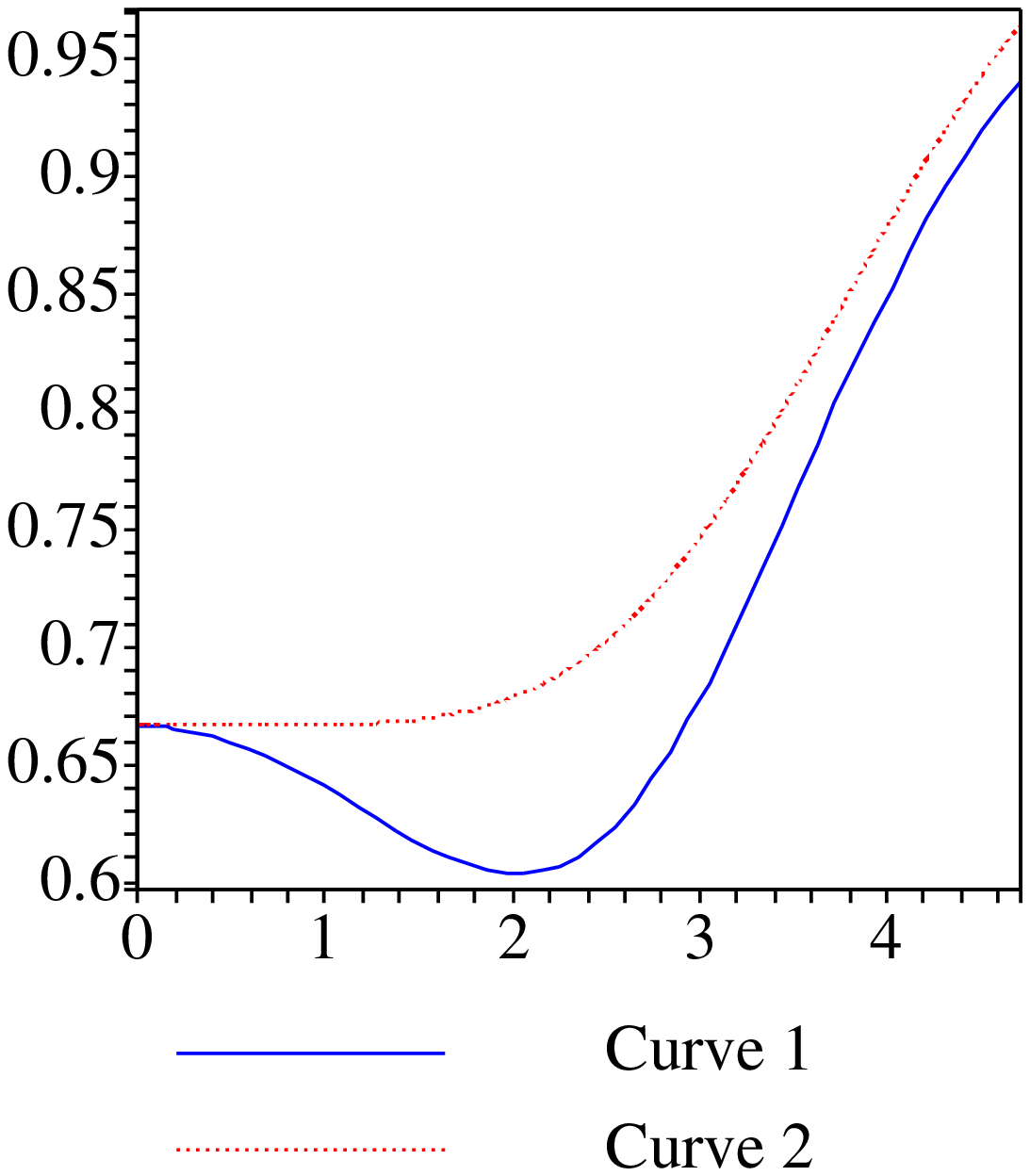}}
\begin{center} {\bf (b)} \end{center}
\end{minipage}
\hfill
\caption{The evolution of the toy model as a function of the
development angle $\phi$.
(a): The deceleration parameter $q$ in the toy model (blue, solid)
and in the \LCDM model (red, dash-dot).
(b): The Hubble parameter multiplied by time, $Ht$, in the toy model
(blue, solid) and in the \LCDM model (red, dash-dot).}
\label{fig:toy}
\end{figure}

The toy model has one free parameter, the relative size of the
two regions at some time. For illustration purposes, we
fix this by setting the deceleration parameter
$q\equiv-\addot/a/H^2$ at $\phi=3\pi/2$ to the value
of the spatially flat \LCDM FRW model with $\Omega_\Lambda=0.7$.
In \fig{fig:toy} (a) we plot $q$ as a function
of the development angle $\phi$. In addition to the
toy model, we show the \LCDM model for comparison.

The \LCDM model starts matter-dominated, with $q=1/2$.
As vacuum energy becomes important, the model
decelerates less and then crosses over to acceleration.
Asymptotically, $q$ approaches $-1$ from above as the
Hubble parameter approaches a constant.
The backreaction model also starts with the FRW
matter-dominated behaviour, then the expansion slows
down more, before $q$ turns around and the expansion
decelerates less and eventually accelerates:
in fact the acceleration is stronger than in the \LCDM model.

The acceleration is not due to regions speeding up
locally, but due to the slower region becoming less
represented in the average.
First the overdense region brings down the expansion rate,
but its fraction of the volume falls because of the slower
expansion, so eventually the underdense region takes over
and the average expansion rate rises. This is particularly
easy to understand after the overdense region has started
collapsing at $\phi=\pi$. Then the contribution $v_2 H_2$ 
of the overdense region to \re{Hex} is negative, and
its magnitude shrinks rapidly as $v_2$ decreases,
so it is transparent that the expansion rate increases.
Note that while there is an upper bound on the expansion
rate, there is no lower bound on the collapse rate.
Therefore, the acceleration can be arbitrarily rapid,
and $q$ can even reach minus infinity in a finite time.
(This simply means that the collapsing region
becomes so dominant that $H^2$ vanishes in the denominator of $q$.)
This is in contrast to FRW models, where $q\geq-1$
unless the null energy condition (or the modified gravity
equivalent) is violated.
After the overdense region stops being important, the expansion
rate will be given by the underdense region alone, and the
expansion will again decelerate. Acceleration is a transient
phenomenon associated with the volume becoming dominated
by the underdense region.

Figure \ref{fig:toy} (b) shows the Hubble parameter
multiplied by time as a function of the development angle
$\phi$. This contains the same information as \fig{fig:toy} (a),
but plotted in terms of the first derivative of the scale
factor instead of the second derivative.
In the \LCDM model, $Ht$ starts from $2/3$ in the
matter-dominated era and increases monotonically without
bound as $H$ approaches a constant. In the toy model,
$Ht$ falls as the overdense region slows down, then rises
as the underdense region takes over, approaching unity
from below. The Hubble parameter in the toy model is
smaller than in the \LCDM model at all times, and because
$H$ is bounded from above by the fastest local expansion rate,
$Ht$ cannot exceed unity. This bound also holds in realistic models:
as long as the matter can be treated as dust and vorticity
can be neglected, we have $Ht\leq1$ at all times \cite{Rasanen:2005},
in contrast to FRW models with exotic matter or modified gravity.
This is a prediction of backreaction.
(For discussion of vorticity and non-dust terms
in the energy-momentum tensor, see \cite{Rasanen:2009b, Buchert:2001}.)

Whether the expansion accelerates depends on how rapidly the
faster expanding regions catch up with the slower ones, roughly
speaking by how steeply the $Ht$ curve rises.
This is why the variance contributes positively to acceleration:
the larger the variance, the bigger the difference between
fast and slow regions, and the more rapidly the fast regions
take over.

\section{Towards reality}

\subsection{A statistical semi-realistic model}

The toy model shows how acceleration due to inhomogeneities
can occur and makes transparent what this means physically.
Acceleration has also been demonstrated with the exact spherically
symmetric dust solution, the Lema\^{\i}tre-Tolman-Bondi model
\cite{Chuang:2005, Paranjape:2006, Kai:2006}.
So there is no ambiguity: accelerated average expansion due to
inhomogeneities is possible.
The question is whether the distribution of structures in the
universe is such that this mechanism is realised.
The statement that faster expanding regions increase
their volume more rapidly makes it sound as if there would
necessarily be less deceleration (if not acceleration)
than in the FRW case. For a set of isolated regions, this is
true: eventually, the volume will be dominated by the fastest region.
However, the characteristic feature of structures in the real
universe is their hierarchical buildup. Smaller
structures become incorporated into larger ones,
and rapidly expanding voids can be extinguished by
collapsing clouds.

The non-linear evolution of structures is too complex to
follow exactly. However, because the
universe is statistically homogeneous and isotropic,
statistical properties are enough to evaluate
the average expansion rate.
In terms of the Buchert equations \re{Ray}--\re{cons},
the average expansion rate is completely determined
by the variance and the average shear scalar.
If one wants an upper limit on the acceleration, discarding
the shear and looking only at the variance would be enough.
The average expansion rate is also determined if we
know which fraction of the universe is in which
state of expansion or collapse.
Instead of trying to find a solution for the metric
and calculating the quantities of interest from it,
it is useful to consider an ensemble of regions from which
we can determine the average expansion rate without having to
consider the global metric.
We now discuss a semi-realistic model which
does this by extending the two fixed regions of the
toy model to a continuous distribution of regions which evolves
in time \cite{Rasanen:2008a, peakrevs}.

The starting point is the spatially flat matter-dominated FRW
model with a linear Gaussian field of density fluctuations.
Structure formation, even though complicated, is a deterministic
process. Therefore any statistical quantity at late times
is determined by the initial distribution processed by
gravity. For a Gaussian distribution, the power spectrum contains
all information. So even in the completely non-linear regime, the average
expansion rate follows from the power spectrum. The problem
is formulating a tractable model for propagating the structures given
by the initial power spectrum into the non-linear regime with gravity.
One approach, proposed in \cite{Bardeen:1986}, is to identify
structures at late times with spherical peaks in the original
linear density field, smoothed on the appropriate scale.
The number density of peaks as a function of the smoothing
scale and peak height can be determined analytically in terms
of the power spectrum.
In the original application, the correspondence between peaks
and structures was assumed to hold only for very non-linear
overdense structures: all peaks exceeding a certain
density threshold were identified with stabilised structures.
Here the idea is a bit different: spherical peaks of any
density are identified with structures having the same linear
density contrast. Troughs are identified with spherical
voids in the same way.
(As the distribution is Gaussian, the statistics of peaks
and troughs are identical.)
We keep the smoothing threshold fixed such
that $\sigma(t,R)=1$, where $\sigma$ is the root mean square
linear density contrast, $t$ is time and $R$ is the smoothing scale.
Non-linear structures form at $\sigma\approx1$, so
$R$ corresponds to the size of the typical largest structures,
and grows in time. The smoothing is just a simplified
treatment of the complex stabilisation and evolution of
structures in the process of hierarchical structure formation.

Since the peaks are spherical and isolated, and they are individually
assumed to be in the Newtonian regime, their expansion rate
is the same as that of a dust FRW universe with the same density,
as in the toy model. The fraction of volume which is neither in
peaks nor in troughs is taken to expand like the spatially
flat matter-dominated FRW model.

\begin{figure}[t]
\centering
\scalebox{0.5}
{\includegraphics[angle=0, clip=true, trim=0cm 0cm 0cm 0cm, width=\textwidth]{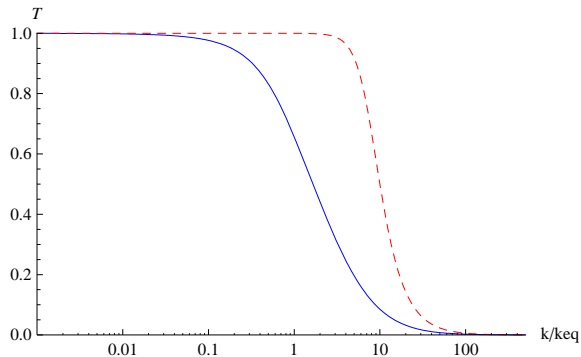}}
\caption{The BBKS (blue, solid) and BDG (red, dashed) transfer functions as a function of $k/\keq$.}
\label{fig:transfer}
\end{figure}

The peak number density as a function of time is
determined by the power spectrum, which consists of two
parts: the primordial power spectrum,
determined in the early universe by inflation or some
other process, and the transfer function, which describes the
evolution between the primordial era and the time when the
modes enter the non-linear regime. The transfer function
$T(k)$ simply multiplies the amplitude of the primordial modes.
We take a scale-invariant primordial spectrum
with the amplitude determined from the CMB anisotropies;
small variations from scale-invariance have little effect.
For the transfer function, we assume that the dark
matter is cold, and we consider two different
approximations in order to show the uncertainty
in the calculation. The BBKS transfer function \cite{Bardeen:1986}
is a fit to numerical calculations (we take a baryon fraction
of $0.2$), and the BDG form introduced in \cite{Bonvin:2006}
is a simple analytically tractable function
with the correct qualitative features.
The transfer functions are shown in \fig{fig:transfer},
as a function of $k/\keq$, the wavenumber divided by the
matter-radiation equality scale.
Modes with $k>\keq$ enter the horizon during radiation domination,
so their amplitude is damped, and the sooner they enter, the
more they are damped before the universe becomes matter-dominated,
so there is an (approximately) $k^{-2}$ damping tail.
Modes with $k<\keq$ enter during the matter-dominated era
and retain their original amplitude, and for modes with $k\sim\keq$,
the transfer function interpolates between these two regimes.
In the BBKS transfer function, the transition is centered around
$\keq$ and is more smooth, while in the BDG case the transition
happens a bit earlier and is more rapid. (Even the more
realistic BBKS transfer function has an error of 20--30\%
compared to Boltzmann codes.)

We have
\bea \label{H}
  H(t) = \int_{-\infty}^{\infty} \rmd\delta\, v_\delta(t) H_\delta(t) \ , 
\eea

\noindent where $\rmd\delta v_\delta$ is the fraction of
volume in regions with linear density contrast $\delta$
and expansion rate $H_\delta(t)$.
The correspondence between $\delta$ and $H_\delta$ is given
by the spherical evolution model (i.e. FRW evolution),
and the distribution of regions $v_\delta(t)$ is given by the
peak statistics, which is determined by the power spectrum
of the Gaussian density field. 
With the transfer function fixed, there are no free
parameters: the expansion history $H(t)$ given by \re{H}
is completely determined.
Since the primordial spectrum is scale-invariant and the smoothing
and peak identification process does not introduce a scale,
features in the expansion rate as a function of time can only come
from the turnover at the matter-radiation equality scale in the transfer
function.

\begin{figure}
\hfill
\begin{minipage}[h]{6cm} 
\scalebox{1.0}{\includegraphics[angle=0, clip=true, trim=0cm 0cm 0cm 0cm, width=\textwidth]{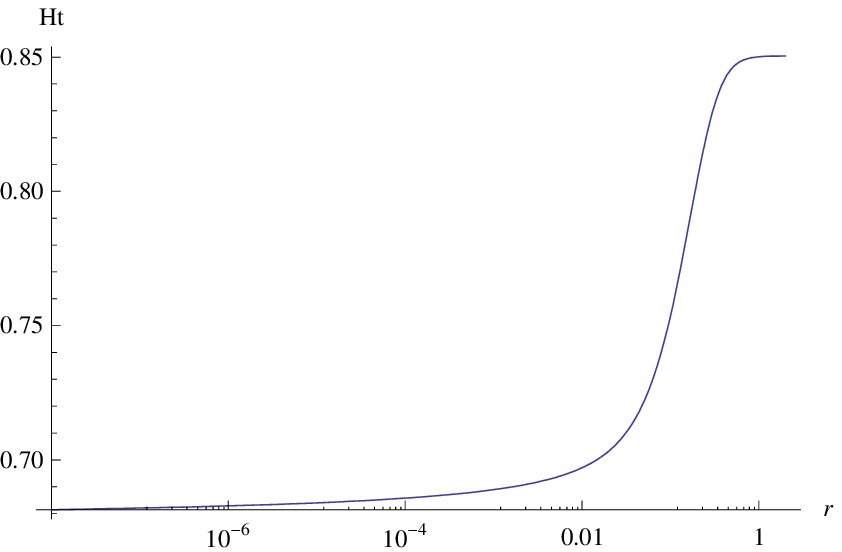}}
\begin{center} {\bf (a)} \end{center}
\end{minipage}
\hfill
\begin{minipage}[h]{6cm}
\scalebox{1.0}{\includegraphics[angle=0, clip=true, trim=0cm 0cm 0cm 0cm, width=\textwidth]{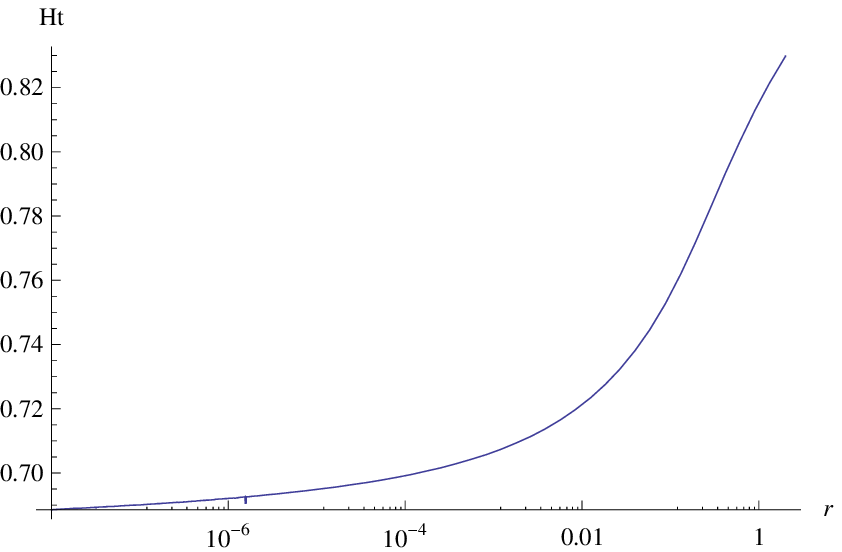}}
\begin{center} {\bf (b)} \end{center}
\end{minipage}
\hfill
\caption{The expansion rate $Ht$ as a function of $r=\keq R$ for
(a) the BDG transfer function and (b) the BBKS transfer function.}
\label{fig:Htr}
\end{figure}

In \fig{fig:Htr} we show $Ht$ as a function of
$r\equiv\keq R$, the smoothing scale relative to the
matter-radiation equality scale. Essentially, the coordinate $r$
is time as measured by the size of the largest generation of structures.
We have $Ht\approx 2/3$ at early times, as the fraction of
volume in non-linear structures is small. As time goes
on, deeper non-linear structures form, and they
take up a larger fraction of the volume.
The expansion rate grows (relative to the FRW value) slowly,
until there is a rapid rise and saturation, roughly at the scale
of matter-radiation equality.
It is clear that after $r=1$, when the perturbations which
correspond to the matter-radiation equality scale collapse,
$Ht$ must settle to a constant, since the transfer function
is nearly unity, and there is no scale in the system anymore.

The matter-radiation equality scale is
$\keq^{-1}\approx 13.7\om^{-1}$ Mpc $\approx$ 100 Mpc,
using the value $\om=0.14$ \cite{Vonlanthen:2010}.
Observationally, $\sigma(t,R)\approx1$
today on scales somewhat smaller than 8 $h^{-1}$ Mpc, so
$R_0\approx$ 10 Mpc. Therefore the present day happens to be
located around $r=0.1$ in the plots -- right in the transition region.
Note that nothing related to present day has been used as input in
the calculation.

\begin{figure}
\hfill
\begin{minipage}[t]{6cm} 
\scalebox{1.0}{\includegraphics[angle=0, clip=true, trim=0cm 0cm 0cm 0cm, width=\textwidth]{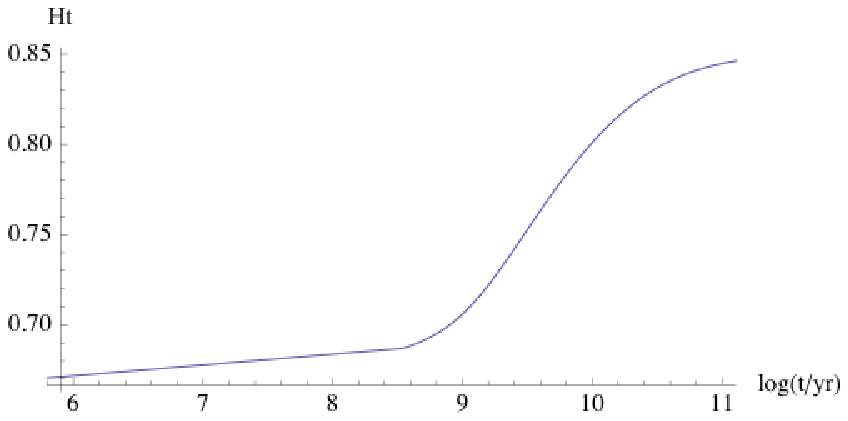}}
\begin{center} {\bf (a)} \end{center}
\end{minipage}
\hfill
\begin{minipage}[t]{6cm}
\scalebox{1.0}{\includegraphics[angle=0, clip=true, trim=0cm 0cm 0cm 0cm, width=\textwidth]{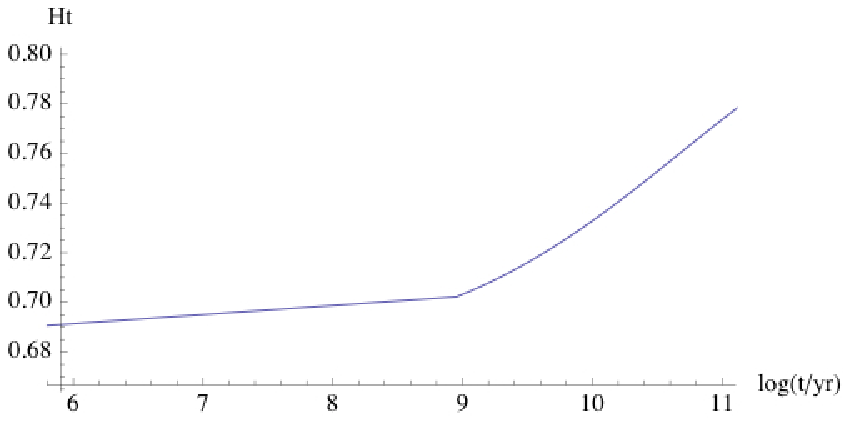}}
\begin{center} {\bf (b)} \end{center}
\end{minipage}
\hfill
\caption{The expansion rate $Ht$ as a function of time for
(a) the BDG transfer function and (b) the BBKS transfer function.}
\label{fig:Ht}
\end{figure}

It is instructive to view $Ht$ also as function of time as measured
in years. In \fig{fig:Ht}, the horizontal axis is $\log_{10}(t/$yr).
For the BDG transfer function, $Ht$ has the FRW value at
one million years, and it grows very slowly until about a billion
years, when $Ht$ starts to rise, and then saturates to a value somewhat
larger than $0.8$ at some tens of billions of years. For the
more realistic BBKS transfer function, the behaviour is
qualitatively the same, but the transition is slower
and the final value of $Ht$ is smaller.
The slope of the $Ht$ curve is less steep as a function
of time than as a function of $r$, because the size of
structures grows more slowly at late times.
When plotting $Ht$ as function of the smoothing
scale, the comparison scale is $\keq$, whereas here it is the time
of matter-radiation equality, $\teq$. Now the amplitude of the
primordial perturbations also enters.
The timescale follows from the shape of the transfer function.
Perturbations which entered the horizon at matter radiation equality reach
non-linearity at $t\approx A^{-3/2}\teq\approx100$ Gyr\footnote{We have
$1=\sigma(t,R=\keq^{-1})=\left( \int_0^{\keq}\frac{\rmd k}{k} \frac{4}{9} \frac{k^4}{(a H)^4} \Delta_\Phi(k) T(k)^2 \right)^{\ha} = \frac{A}{3} \frac{(\aeq H_\mathrm{eq})^2}{(aH)^2} = \frac{A}{3} \frac{t^{2/3}}{\teq^{2/3}}$,
where $\Delta_\Phi(k)=A^2$, and we have adopted the step function
as the window function and approximated $T(k)=1$ for $k\leq\keq$.},
where $A=3\times10^{-5}$
is the primordial amplitude and the matter-radiation equality
time is $\teq\approx 1000 \om^{-2}$ years $\approx$ 50 000 years
for $\om=0.14$. This is when the expansion rate saturates, and
it enters the transition region somewhat earlier.

As noted in \sec{sec:toy}, whether or not the expansion
accelerates is a quantitative question
related to the slope of the $Ht$ curve.
In the present case, while the expansion rate increases relative
to the FRW value, the change is not sufficiently rapid for the
expansion to accelerate, there is just less deceleration.
This is related to the fact that, unlike in the toy model,
the overdense regions play almost no role, and
the evolution of $Ht$ can be understood in
terms of the underdense voids.
At early times, voids take up only a small part of the volume,
and $Ht$ rises smoothly as their volume fraction increases.
In order to obtain a more drastic change in $Ht$, the expansion
rate should have extra deceleration due to overdense regions
before the voids take over, as in the toy model.
(This effect is present in the model, but it is too small
to be visible in \fig{fig:Htr} and \fig{fig:Ht}.)
If the expansion were to slow down more before the voids take over,
the variance and the change in the expansion rate would be larger.
The magnitude of the change of $Ht$ is easy to
understand: if the universe were completely dominated by completely
empty voids, we would have $Ht=1$. Since not all of the volume is
taken up by voids and they are not totally empty, $Ht$ is
somewhat smaller than unity.

It is encouraging that the model gives a change of the right order of
magnitude in $Ht$, 15--25\%, and that the timescale for the change
comes out right. However, the model involves many approximations, such as
treating structures as spherical, using an approximate transfer
function, having an artificial split between the peaks/troughs
and the smooth space, not taking into account that the Gaussian symmetry
between the overdense and underdense regions is broken in the non-linear
regime (in the present treatment, they have equal mass in at all times) and
treating the structures as isolated even for small density contrasts
and high peak number densities.
It is clear the model cannot be trusted beyond an order of magnitude.
It is also possible that a more careful statistical treatment would
reveal some cancellations that would significantly reduce the effect
of backreaction from this approximate estimate.

\subsection{Beyond Newton}

A detailed statistical treatment would require significantly
more work than put into the simple semi-realistic model.
Before investing the effort, one may ask why bother, since
N-body simulations provide detailed information about the
non-linear regime of structure formation. The problem is that
the simulations use Newtonian gravity with periodic boundary
conditions. In Newtonian gravity, the variance and the shear cancel in
the backreaction variable $\sQ$ given in \re{Q}, up to total derivatives
which can be written as boundary terms \cite{Buchert:1995}.
Boundary terms vanish for periodic boundary conditions, but using a large
simulation and considering boxes of the size of the observable universe
would not help the situation. Total derivative terms represent a flux,
and due to statistical homogeneity and isotropy, the integrated flux
over the whole boundary should vanish (up to statistical fluctuations),
as otherwise there would be a preferred direction.

In general relativity, the backreaction variable
$\sQ$ does not reduce to a boundary term,
and the average expansion rate of a volume depends on the behaviour
everywhere in the volume, not just on the boundary.
In contrast, the Newtonian evolution is sensitive to boundary
conditions, even for infinitely far away boundaries, which
is related to the fact that Newtonian cosmology does
not have a well-posed initial value problem.
This is one aspect of the qualitative difference between general
relativity and what is called Newtonian cosmology.
The small-velocity, weak field limit of general relativity is not
Newtonian gravity, as demonstrated by the existence of Newtonian
solutions which are not the limit of any general relativity solution
\cite{Ellis:1971, Senovilla:1997}.
Rather, it is a theory with new degrees of freedom and
additional constraints compared to Newtonian gravity
\cite{Ellis:1971, Ellis:1994, Ehlers:1999, Szekeres:2000, Rasanen:2010a}.
The formulation of this limit of general relativity in the
cosmological setting with non-linear perturbations is an open issue.

It is important to make sure that an improved statistical
treatment would be consistent with the relativistic equations of
motion and constraints.
One may ask why the average expansion rate in the peak model
differs from the FRW result.
After all, the individual regions were taken to be Newtonian,
and there is at first sight nothing non-Newtonian
about the identification of the peaks as structures.
In Newtonian gravity, the feature that inhomogeneities do not change
the average expansion rate in a statistically homogeneous and isotropic
universe can be understood from energy conservation.
In the exactly homogeneous and isotropic case, the Newtonian Friedmann
equation (multiplied by $a^2$) can be interpreted as stating
that the kinetic energy plus the potential energy is constant.
The relativistic Friedmann equation is mathematically identical,
but has a different physical interpretation, with the
constant energy replaced by the spatial curvature term.
However, the correspondence does not hold beyond the FRW case.
In Newtonian gravity, the total energy is conserved even
when the system is inhomogeneous and anisotropic, as long
as the force is conservative and the system is isolated
(i.e. the boundary terms in $\sQ$ vanish).
However, in general relativity, there is no conservation
law for the average spatial curvature, and
$a^2\av{\sR}$ is in general not constant.
In the peak model, the average spatial curvature evolves from zero
to large negative values as the volume becomes dominated by
underdense voids. This is possible in general relativity, while
in Newtonian gravity the total energy cannot change.

\section{Light propagation} \label{sec:light}

\subsection{The redshift}

Let us say we were to have a reliable calculation of the
average expansion rate, either via an improved statistical
model or by including the relevant general relativistic
degrees of freedom in a simulation. What would this tell us
about observations? The average is taken on a spacelike
hypersurface of simultaneity, but we can only observe things
inside the past lightcone. Most cosmological observations
are made along the lightcone, measuring the redshift and the
angular diameter (or luminosity) distance.
In a general spacetime, these quantities are not determined
solely by expansion, and certainly not by the average
expansion rate along spacelike slices.
However, in a statistically homogeneous and isotropic universe
where the distribution evolves slowly, the average expansion
rate does give the leading behaviour of the redshift and the
distance \cite{Rasanen:2008b, Rasanen:2009b}.
In a general dust spacetime, the redshift is given by
\bea \label{z}
  1 + z &=& \exp\left( \int_{\eta}^{\eta_0} \rmd \eta \left[ \frac{1}{3} \theta + \sigma_{\a\b} e^\a e^\b \right] \right) \ ,
\eea

\noindent where $\eta$ is defined by
$\pat/\pat\eta \equiv (u^\a + e^\a) \pat_\a$, and $e^\a$
is the spatial direction of the null geodesic. If there are no
preferred directions and the change in the distribution
is slow compared to the time it takes for a light ray to
pass through a homogeneity scale sized region, the integral
over $\sigma_{\a\b} e^\a e^\b$ is suppressed.
In the real universe, the homogeneity scale of around 100 Mpc
is indeed much smaller than the timescale for the change in
the distribution, which is given by the Hubble scale
$H_0^{-1}=3000 h^{-1}$Mpc. In the early universe, structure formation
was less advanced, so further down the null geodesic the
homogeneity scale is even smaller relative to the Hubble scale.
The direction $e^\a$ changes slowly for typical light rays
\cite{Rasanen:2009b}, whereas the dust shear is correlated with
the shape and orientation of structures, and changes on the
length scale of those structures. If there are no preferred
directions, structures are oriented in all
directions equally over large scales,
so $\sigma_{\a\b}$ should contribute via its
trace, which is zero. Therefore the integral over
$\sigma_{\a\b} e^\a e^\b$ should vanish, up to statistical
fluctuations and corrections from correlations between 
$\sigma_{\a\b}$ and $e^\a$ and evolution of the distribution.
We can split the local expansion rate as $\theta=\av{\theta}+\Delta\theta$,
where $\Delta\theta$ is the local deviation from the average, and
similarly argue that the integral of $\Delta\theta$ is suppressed
relative to the contribution of the average expansion rate.

At this point, the choice of hypersurface is important.
For the cancellations to occur, the hypersurface of
averaging has to be the hypersurface of statistical homogeneity
and isotropy. (In addition, the evolution of the distribution
from one hypersurface to another has to be slow compared to
the homogeneity scale.)
This defines the hypersurface of averaging:
the primary quantities are the observable redshift and distance,
and averages are useful only insofar as they
give an approximate description of what is observed
\cite{Rasanen:2008b, Rasanen:2009b}.
We have taken the averages on the hypersurfaces of constant
proper time of observers comoving with the matter.
Since the evolution of structures is governed by the proper
time, one can argue that this is close to the hypersurface
of statistical homogeneity and isotropy
\cite{Rasanen:2006b, Rasanen:2008a, Rasanen:2008b}.
The details are likely to be more complicated, but
non-relativistic changes in the four-velocity which
defines the hypersurface lead only to small changes
in the averages \cite{Rasanen:2009b}.

Given that $\av{\theta}=3\adot/a$, we obtain $1+z\approx a(t)^{-1}$,
the same relation between expansion and
redshift as in the FRW case. Note that this result depends
on the fact that the shear and the expansion rate enter into
the integral \re{z} along the null geodesic linearly.
In the case of the equations of motion \re{Rayloc}--\re{consloc}
for the geometry, the shear and the expansion rate
enter quadratically, so the variations do not cancel in the average,
and instead we have the generally non-zero backreaction variable $\sQ$.

\subsection{The distance}

For the angular diameter distance, we can apply similar
qualitative arguments to obtain the result \cite{Rasanen:2008b}
\bea \label{DA}
  H \pat_{\bz} \left[ (1+\bz)^2 H \pat_{\bz} \bar{D}_A \right] &\approx& - 4\pi\GN \av{\rho} \bar{D}_A \ ,
\eea

\noindent where $\bar{D}_A$ is the dominant
part of the angular diameter distance with the corrections to the
mean dropped, and the same for the redshift, $1+\bz\equiv a(t)^{-1}$.
From the conservation of mass, \re{cons}, it follows that
$\av{\rho}\propto(1+z)^3$. The distance is therefore determined
entirely by the average expansion rate $H(z)$ and the normalisation
of the density today, i.e. $\Omega_{\mathrm{m0}}$.
For a general FRW model, $\av{\rho}$ in \re{DA}
would be replaced by $\rho+p$.
So the equation for the mean angular diameter distance in
terms of $H(z)$ in a statistically homogeneous and isotropic
dust universe (with a slowly evolving distribution) is
the same as in the FRW $\Lambda$CDM model.
If backreaction were to produce exactly the same expansion
history as the $\Lambda$CDM model, the distance-redshift
relation would therefore also be identical.
This is the case even though the spatial curvature would
be large\footnote{From \re{Ray} and \re{Ham} we see that
for $\Omega_{\mathrm{m0}}=0.3$ and $q_0=-0.55$,
corresponding to the spatially flat $\Lambda$CDM model
with $\Omega_{\mathrm{\Lambda0}}=0.7$,
we have $\av{\sR}_0=-6.3 H_0^2$, or
$\Omega_{R0}\equiv-\av{\sR}_0/(6 H_0^2)=1.05>1$.
The physical reason for the large spatial curvature
is that most of the volume is occupied by very underdense
regions.}, as the spatial curvature affects the distances
differently than in the FRW case.

Note that in a general spacetime, the luminosity distance is related
to the angular diameter distance by $D_L=(1+z)^2 D_A$
\cite{Ellis:1971}, so (from the theoretical point of view)
it measures the same thing.

Backreaction is not expected to produce an expansion history
identical to the $\Lambda$CDM model: if the expansion accelerates
strongly, then this is likely to be preceded by extra deceleration.
Therefore the distances will also be different.
However, the backreaction distance-redshift relation will be
biased towards the $\Lambda$CDM model, compared to a FRW model
with the same expansion history as in the backreaction case.
The reason is that in the FRW model, the
equation for $D_A$ is modified not only by the
change in $H(z)$, but also by the change in $\rho+p$.
This may help to explain why distance observations prefer
the value $-1$ for the effective equation of state.

It has been pointed out that the relation between $D_A(z)$ and $H(z)$
can be used as a general test of FRW models \cite{Clarkson:2007b}.
If we measure the distance and the expansion rate independently,
we can check whether they satisfy the FRW relation. If they do not,
the observations cannot be explained in terms of any four-dimensional
FRW model.
This holds independent of the presence of dark energy or
modified gravity, because light propagation depends directly
on the geometry of spacetime, regardless of the equations of
motion which determine it.
Similarly, we can test the backreaction conjecture that the
change in the expansion rate at small redshift is due
to structure formation without having a prediction for how
the expansion rate changes, simply by checking whether the
measured $D_A(z)$ and $H(z)$ satisfy \re{DA}.
The relation \re{DA} and the violation of the FRW consistency
condition between expansion and distance is a unique prediction
of backreaction which distinguishes it from FRW models.
However, the derivation of the relation between $D_A(z)$ and $H(z)$
should be done more rigorously, and the expected magnitude of the
violation is unclear.

\section{Summary}

Observations of the universe at late times are inconsistent
with homogeneous and isotropic FRW models which have ordinary
matter and gravity.
The problem is usually addressed by adding exotic matter
or modifying general relativity.
However, non-linear structures also influence the expansion rate:
this is an effect which is present in reality but missing in FRW models.
The Buchert equations which do include the effect of structures
show that it is possible for the average expansion of a clumpy dust
universe to accelerate, and there are toy models which demonstrate this.
The physical explanation is simple: faster expanding
regions increase their fraction of the volume more rapidly,
so the average expansion rate increases.
In a semi-realistic model, the correct timescale
of about $10^5\teq\sim$ 10 billion years and the right order
of magnitude for the change of the expansion rate emerge
from the physics of structure formation without new parameters.

In Newtonian cosmology, backreaction is necessarily small for a
statistically homogeneous and isotropic distribution, but this is
not the case in general relativity. Therefore, if backreaction
has a significant effect on the expansion rate in the real universe,
this is due to non-Newtonian aspects of general relativity.

Even if backreaction is important, the relation between the
redshift and the average expansion rate is the same as in FRW models,
if the distribution of structures is statistically homogeneous and
isotropic and evolves slowly. In contrast, the relation between
the average expansion rate and the angular diameter distance is
different from the FRW case.
This relation is a unique backreaction prediction which makes it
possible to distinguish the effect of non-linear structures from
FRW dark energy or modified gravity models.

The present estimates of the effect of structure formation
on the average expansion rate cannot be trusted beyond
an order of magnitude, and it is possible that a careful
study will reveal cancellations which lead to a negligible effect.
The relation between the average expansion rate
and light propagation should also be studied more rigorously,
and the difference between general relativity and Newtonian
gravity in the cosmological non-linear regime remains to be
fully understood. There is much work to be done before we can
say whether or not the backreaction conjecture that the failure of
ordinary homogeneous and isotropic models at late times
is due to the breakdown of homogeneity and isotropy is correct.
Until this effect has been quantified, we do not know whether
new physics is needed to explain the observations, or if they
can be understood in terms of a complex realisation of the
physics we already know.


\end{document}